\documentclass{PoS}

\newcommand{\tprime}{\ensuremath{t^\prime}}
\newcommand{\sht}{\ensuremath{H_T}}
\newcommand{\met}{\ensuremath{\,/\!\!\!\!E_{T}}}
\newcommand{\pt}{\ensuremath{p_T}}

\newcommand{\tprimebar}{\ensuremath{\bar{t}^\prime}}
\newcommand{\bprime}{\ensuremath{b^\prime}}

\newcommand{\mreco}{\ensuremath{M_{reco}}}

\title{Searches for Massive \tprime quarks decaying to W + q at the Tevatron}

\ShortTitle{Searches for Massive \tprime quarks decaying to W + q at the Tevatron}

\author{\speaker{Alison LISTER}%
        \thanks{On behalf of the CDF and D0 Collaborations}\\
       University of Geneva\\
       E-mail: \email{alison.lister@cern.ch}}

\abstract{We present searches for pair-production of fourth generation \tprime quarks in their decays to $Wq$.
We analyze 4.6~fb$^{-1}$ and 4.3~$fb^{-1}$ of data collected by the CDF and D0 detectors, respectively, at the Fermilab Tevatron collider at a centre-of-mass energy of $\sqrt{s} = 1.96$~TeV.
        	We reconstruct the mass of the heavy quark
          and perform a two-dimensional-fit to the observed (\sht ,\mreco)
          distribution to discriminate the new physics signal
          from standard model backgrounds. 
	  As no significant excess of events is observed, we exclude a fourth-generation 
	  $\tprime$ quark with a mass below 335 ( 296)~GeV at 95\%CL at CDF ( D0).
}

\FullConference{35th International Conference of High Energy Physics - ICHEP2010,\\
		July 22-28, 2010\\
		Paris France}

\begin{document}

\section{Introduction}

These conference proceedings report on preliminary results from CDF~\cite{cdf_note} and D0~\cite{d0_note} investigating
whether the present data allow or preclude
the production of hypothetical new quarks which decay to final states
with a high transverse momentum ($\pt$) lepton, large missing transverse energy ($\met$), 
and multiple hadronic jets,
having large total transverse energy ($\sht$), and thus mimicking
top quark pair event signatures in the lepton+jets decay channel.
Previous versions of this search were conducted at CDF at lower integrated luminosities~\cite{prev,prev_paper}.

We refer to the hypothetical new quark as $\tprime$ for brevity, 
although such a signature could be a standard fourth-generation up-type heavy 
quark in which the splitting between the \tprime mass and the \bprime mass is less 
than the mass of the W boson (so that the decay is predominantly to $Wq$), as 
well as any up-type quark.
For the purpose of setting limits we assume that the new heavy quarks are produced strongly, are heavier than the top quark and decay promptly to $Wq$ final states.

Due to the variety of
theoretical models predicting similar signatures as well as the number of
free parameters within each model, an a priori method was established to estimate the significance of a potential excess of events without
attributing the excess to a particular new physics model.
In our analysis no significant excess is observed; we thus set a limit on the fourth-generation
$\tprime$ quark pair production cross section (times branching ratio of
$\tprime\to Wq$). 

\section{Theoretical Motivation}
There are a number of new physics models predicting heavy quarks with masses above the one of the top quark whose decays produce event signatures similar to those from top quark decays.
The interested reader is kindly asked to refer to the full conference notes and reference therein for further details and list of references~\cite{cdf_note, d0_note}. 
Here we only outline a few cases of interest to motivate this search.

One of the simplest extensions to the standard model with three generations is a fourth chiral generation of massive fermions. Although not popular historically, such an extension is in good agreement with electroweak precision data.
To avoid the $Z -> \nu \bar{\nu}$ constraint from LEP I a fourth generation of neutrinos must be heavier than $1/2$ the mass of the $Z$-boson. Similarly, to avoid the LEP II bounds, a fourth generation of charged leptons must have $m > 101$ GeV.
At the other end of the spectrum, sizable radiative corrections means masses of fourth generation fermions cannot be much higher than the current lower bounds; the masses of the new heavy quarks \tprime\ and \bprime\ should thus be in the range of a few hundred GeV, within the reach of the Tevatron collider.
In most models a small mass splitting between the \tprime\ and \bprime\ is preferred such that the \tprime\ decays predominantly to $Wq$ where the $q$ is one of the known standard model down-type quarks.
The present bounds on the Higgs are relaxed in the presence of a four-generation model leading to Higgs masses as large as 500 GeV.
Furthermore CP violation is significantly enhanced to a magnitude that might account for the baryon asymmetry of the universe.

\section{Data Samples, Monte Carlo Simulation} 
The data are collected at the CDF and D0 detectors at the Fermilab Tevatron proton-antiproton collider at a centre-of-mass energy of $\sqrt{s} = 1.96$~TeV. 
The CDF and D0 detectors are described elsewhere~\cite{cdf_det, d0_det}.
The latest analysis from CDF (D0) uses 4.6~fb$^{-1}$ (4.3~fb$^{-1}$ ) of integrated luminosity.
The events must have fired one of several trigger conditions, all but one requiring an isolated electron or muon with a high transverse momentum. CDF increased the signal acceptance by incorporating muons collected with a \met\ + jets trigger.

The $\tprime \tprimebar$ signal is generated with {\sc pythia} 6~\cite{py} (v6.216 for CDF and  v6.409 for D0). For the CDF analysis the branching ratios are not modified, for the D0 analysis the branching ratio to $Wb$ is set to 100\%. 
In both cases, as no b-jet identification is applied, the results are applicable to both heavy quarks decaying to a $W$ and light or a heavy quark.
We assume that the \tprime\ is a narrow resonance and as such the intrinsic width will be much smaller than the detector resolution and therefore the exact value of the resonance width does not affect the analysis.
Various mass points were generated for both analyses in order to cover the \tprime\ mass range between 200 and 500 GeV.

The background consists mostly of 
\begin{itemize}
\item $t\bar{t}$ which is modeled with {\sc alpgen}~\cite{alpgen} and use {\sc pythia} for the parton shower for D0 and {\sc pythia} 6.216 for CDF, 
\item $W+jets$ which is modeled with {\sc alpgen} (v2.10$^\prime$ for CDF and v2.11 for D0) and use {\sc pythia} (v6.325) for the parton shower with a jet-matching algorithm following the MLM prescription~\cite{mlm},
\item Multijet events are those events were one of the jets fakes a lepton; this contribution is modeled using data. The methods used by CDF and D0 to obtain the rate and shapes of the QCD background are described in the public notes~\cite{cdf_note}~\cite{d0_note}.
\end{itemize}
Smaller contributions from dibosons, $Z+jets$ and single top are also taken into account.

\section{Event Selection}
While the specifics of the cuts differ between the two experiments, the principles of the cuts are similar.
In addition to the trigger selection mentioned in the previous selection, events are selected by requiring
a single isolated high-\pt\ electron or muon,
high missing transverse energy,
and
four or more jets in the event (with slightly higher cuts on the leading one or two jets). 

In the CDF analysis two additional classes of cuts are applied.
The first set of cuts are on the transverse mass of the $W$ and the missing transverse energy significance to reduce the contribution from QCD events which are notoriously difficult to measure, especially the tails where the signal is expected to be important. Such cute are also implemented in the D0 analysis.
The second set of cuts are targeted at cleaning up the tails of the distributions and to as much as possible remove mis-measured objects causing large values of missing transverse energy.  
These are cuts that are applied to the high-\pt\ electrons, muons and jets and ensure that the missing energy is not back to back with the lepton and not roughly collinear with the jets.
These two sets of additional cuts increased our confidence in the modeling of the tails of the distributions used to fit the \tprime\ signal.
The additional clean-up cuts used by CDF are described in the latest public note~\cite{cdf_note}.


\section{Variable Selection and Fitting}
We use two variables to distinguish the \tprime\ signal from the standard model backgrounds: the reconstructed mass of the top (\tprime) for CDF (D0) (\mreco) and the scalar sum of the transverse momentum of all jets, the charged lepton and the missing transverse energy (\sht). 
Both variables exploit the large mass of the \tprime\ quark.

For both experiments the mass is reconstructed using the minimization of a $\chi^2$ fit describing how compatible the event is with a top (\tprime) decay hypothesis. While the exact form of the $\chi^2$ functions differ, the concept and performance is similar between the two experiments.

The CDF analysis further separates the data into four different sets of distributions: events with exactly 4 jets and those with 5 or more. Each is further separated into events where the $\chi^2$ value of the best fit is less than or greater than 8. The fit is performed simultaneously to all four sets of distributions.

The two-dimensional binned distributions of \sht\ vs \mreco\ is used to test for the presence of a $\tprime \tprimebar$ signal in the data. 
As no signal was observed, the 95\% confidence level upper limits on the production cross section are computed.

The CDF analysis calculates a likelihood as a function of the  signal production cross section and uses Bayes' Theorem to convert it into a posterior density in $\sigma_{\tprime \tprimebar}$. This is then used to set an upper limit or measure the production rate of the signal.
The production rate for W+jets in each jet bin are two free unconstrained independent parameters of the fit. Other parameters such as the $t\bar{t}$ production cross section, lepton ID data/MC scale factors and integrated luminosity are related to systematic errors and treated in the likelihood as nuisance parameters constrained within their expected (normal) distributions. 
We adopt the profiling method for dealing with these parameters, i.e. the likelihood is maximised with respect to the nuisance parameters.
Systematic uncertainties, such as the jet energy scale, the W+jets $Q^2$ scale uncertainty of the initial and final state radiation uncertainty enter the likelihood as gaussian constraint penalty terms. Outside the interval [-1,1] the penalty terms are extrapolated as a linear function of the jet energy scale parameter. This treatment is called vertical template morphing.

The D0 analysis uses as a test statistic the likelihood ratio $L = -2 log \frac{P_{S+B}}{P_B}$, where $P_B$ ($P_{S+B}$) is the Poisson likelihood to observe the data under the background only (signal plus background) hypothesis.
For the background only hypothesis, three components are fit to the data: $t \bar{t}$ production constrained to its cross section, W+jets constrained to the number of expected events with their uncertainty, all other backgrounds added together in their expected proportions and constrained with an overall normalization that floats freely.
To determine the cross section limit, the $CL_S$ method is used~\cite{cls}.
The systematic uncertainties are treated in a similar way as for CDF.

The CDF analysis uses 28 bins for \sht\ and 18 for \mreco\, with the overflow bins defined as events with \sht\ > 800 and \mreco\ > 500 GeV.
Due to limited MC statistics, an algorithm was developed to merge contiguous bins with low total MC statistics to form super-bins used in the likelihood. This procedure by construction deteriorates the sensitivity of the new physics signal, however it eliminates the abnormalities from sources such as due to bins with zero predictions and thus provide reliable observed limits. The exact algorithm and criteria used to define the merging are described elsewhere~\cite{cdf_note}.

\section{Results}

\begin{figure}[htbp]
\begin{center}
\includegraphics[width=0.49\textwidth]{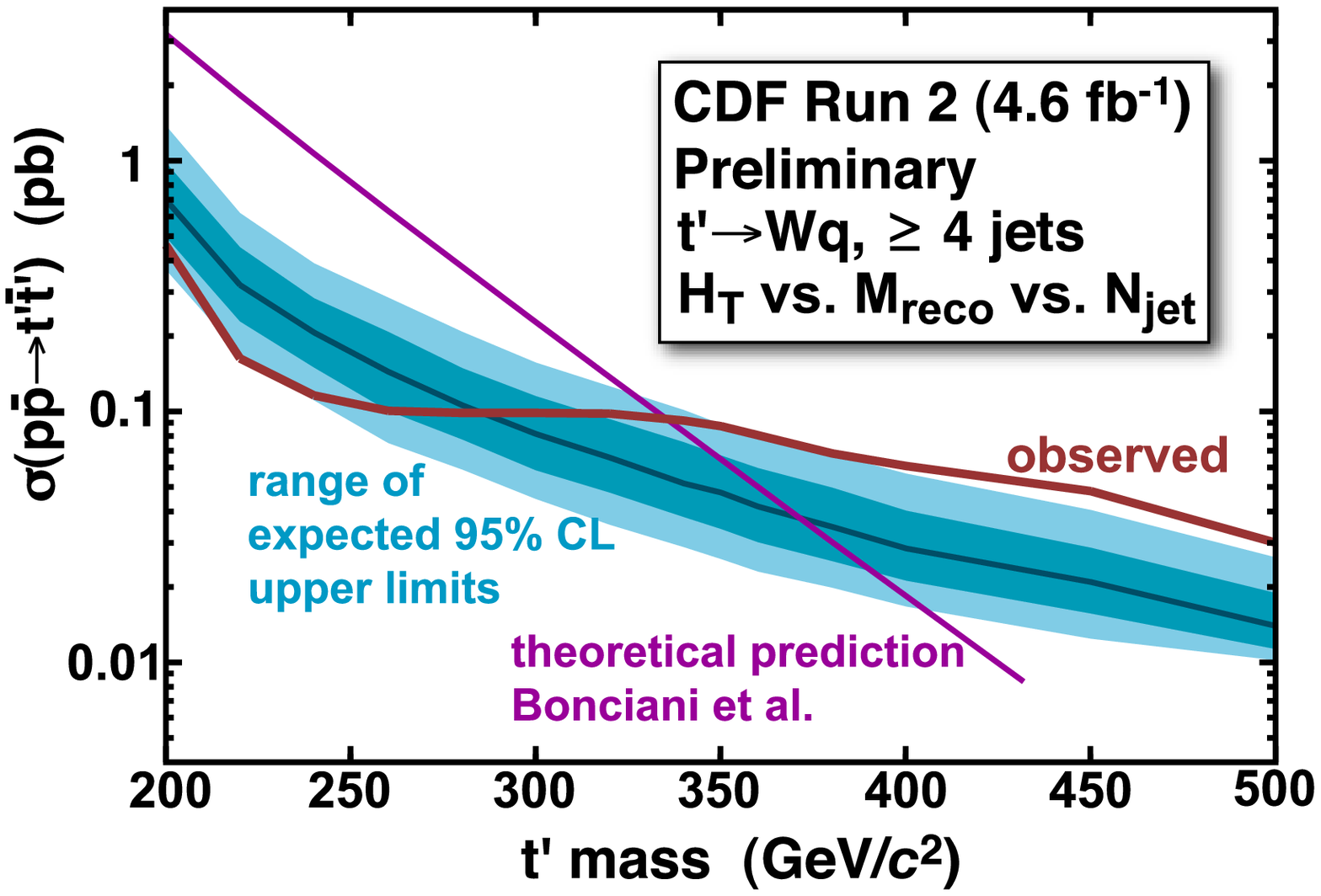}
\includegraphics[width=0.49\textwidth]{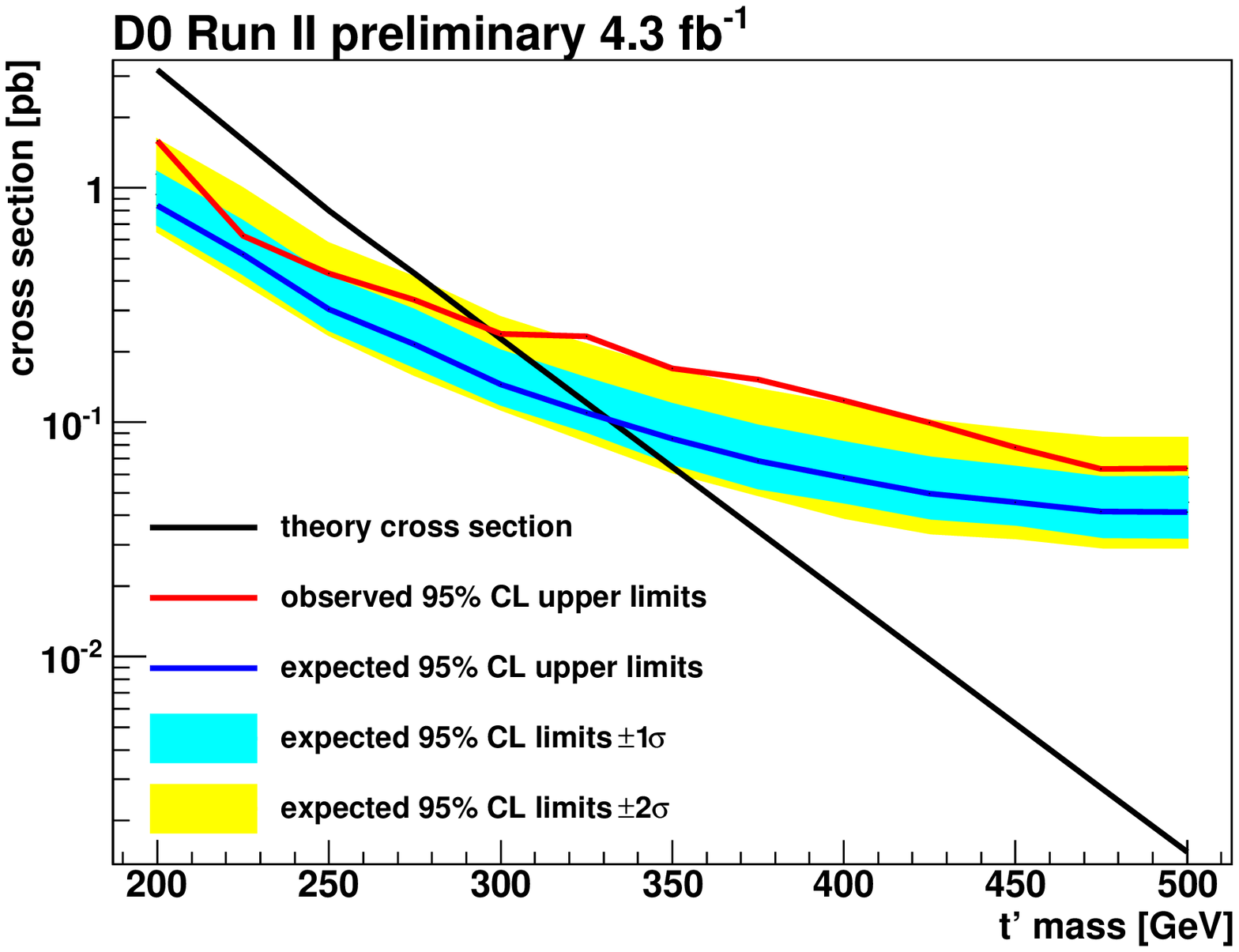}
\end{center}
\vspace{-1cm}
\caption{Observed and expected cross section limits on $\tprime \rightarrow Wq$ compared to theory for CDF (left) and D0 (right). 
}
 \label{fig:results}
\end{figure}

The sensitivity of the methods are tested by drawing pseudo-experiments from standard model distributions. 
The range of expected 95\% CL upper limits within one and two standard deviations are shown in Fig.~\ref{fig:results} as the colored bands. 
The straight line falling curves represent the theoretical predictions. 
The left plot shows the results from CDF, the right one is from D0.
From these figured it follows that given no \tprime\ presence, the methods are on average sensitive to setting upper limit at \tprime\ masses of 372 GeV (CDF) and 330 GeV (D0).
We perform the analysis fit on data and determine the upper limits on the \tprime\ signal. The red curves show the final results.

Based on these results we conclude that there is no significant excess of observed events in either experiment and we can exclude a \tprime\ mass below 335 GeV at CDF and 296 GeV at D0, given the true top mass is 172.5 GeV.

\begin{thebibliography}{99}

  \bibitem{cdf_note} CDF Collaboration, CDF Public note 9846.
\bibitem{d0_note} D0 Collaboration, Conference Note 5892-CONF.


  \bibitem{prev} CDF Collaboration, CDF Public note 7113.
   CDF Collaboration, CDF Public note 7912. 
   CDF Collaboration, CDF Public Note 8495.
  CDF Collaboration, CDF Public Note 9234.

   \bibitem{prev_paper} CDF Collaborataion: T. Aaltonen {\em et al.}, Phys.Rev.Lett.100:161803,2008. [arXiv:hep-ex/08013877].


  \bibitem{cdf_det}
    F. Abe, et al., Nucl. Instrum. Methods Phys. Res. A {\bf 271},
    387 (1988);
    D. Amidei {\em et al.} Nucl. Instum. Methods Phys. Res. A {\bf
    350}, 73 (1994);
    F. Abe {\em et al.}, Phys. Rev. D {\bf 52}, 4784 (1995);
    P. Azzi {\em et al.}, Nucl. Instrum. Methods Phys. Res. A {\bf
    360}, 137 (1995);
    The CDFII Detector Technical Design Report,
    Fermilab-Pub-96/390-E.
    
    \bibitem{d0_det} D0 Collaboration, S. Abachi et al., Nucl. Instrum. Methods Phys. Res. A {\bf 338}, 185 (1994). D0 Collaboration, V.M. Abazov et al., Nucl. Instrum. Methods Phys. Res. A {\bf 565}, 463 (2006).

    \bibitem{py} T. Sj\"ostrand, S. Mrenna, and P. Skands, JHEP 0605:026 (2006).
    
    \bibitem{alpgen} M. L. Mangano {\em et al.}, JHEP 0307:001 (2003), [arXiv:hep-ph/0206293].
    \bibitem{mlm} S. H\"oche {\em et al.}.,  [arXiv:hep-ph/060203].
    

\bibitem{cls} T. Junk, Nucl. Instrum. Methods Phys. Res., A {\bf 434}, 435 (1999).

\end{thebibliography}
\end{document}